# PENERAPAN TEKNIK WEB SCRAPING PADA MESIN PENCARI ARTIKEL ILMIAH

**Ahmat Josi[1), Leon Andretti Abdillah[2), Suryayusra[3)**
[1,3]Program Studi Teknik Informatika, Fakultas Ilmu Komputer, Universitas Bina Darma
[2]Program Studi Sistem Informasi, Fakultas Ilmu Komputer, Universitas Bina Darma
Jl. Ahmad Yani No. 12, Palembang, 30264
Telp : (0711) 515679, Fax : (0711) 515581
E-mail : ahmat_josi@yahoo.com[1), leon.abdillah@yahoo.com[2*), suryayusra@binadarma.ac.id[3)

*Abstract*

*Search engines are a combination of hardware and computer software supplied by a particular company through the website which has been determined. Search engines collect information from the web through bots or web crawlers that crawls the web periodically. The process of retrieval of information from existing websites is called "web scraping." Web scraping is a technique of extracting information from websites. Web scraping is closely related to Web indexing, as for how to develop a web scraping technique that is by first studying the program makers HTML document from the website will be taken to the information in the HTML tag flanking the aim is for information collected after the program makers learn navigation techniques on the website information will be taken to a web application mimicked the scraping that we will create. It should also be noted that the implementation of this writing only scraping involves a free search engine such as: portal garuda, Indonesian scientific journal databases (ISJD), google scholar.*

*Abstrak*

*Search engine yaitu kombinasi perangkat keras dan perangkat lunak komputer yang disediakan oleh perusahaan tertentu melalui website yang telah ditentukan. Search engine mengumpulkan informasi dari web melalui program bot atau web crawler yang secara periodik menelusuri web. Proses pengambilan informasi dari website-website yang ada ini disebut dengan "web scraping". Web Scraping adalah suatu teknik penggalian informasi dari situs web. Web Scraping berkaitan erat dengan pengindeksan web, adapun cara mengembangkan teknik web scraping yaitu dengan cara pertama Pembuat program mempelajari dokumen HTML dari website yang akan diambil informasinya untuk di tag HTML tujuannya ialah untuk mengapit informasi yang diambil setelah itu pembuat program mempelajari teknik navigasi pada website yang akan diambil informasinya untuk ditirukan pada aplikasi web scraping yang akan kita buat. Perlu pula diperhatikan bahwa implementasi scraping pada tulisan ini hanya melibatkan mesin pencari yang gratis seperti: portal garuda, Indonesian scientific journal database (ISJD), google scholar.*

**Kata kunci:** *Pengumpul jaringan, mesin pencari, artikel ilmiah.*

## 1. PENDAHULUAN

Meningkatnya kebutuhan akan informasi mendorong manusia untuk mengembangkan teknologi-teknologi baru agar pengolahan data dan informasi dapat dilakukan dengan mudah dan cepat. Salah satu teknologi yang sedang berkembang dengan pesat saat ini adalah teknologi informasi/komputer (Abdillah & Emigawaty, 2009), teknologi *internet*. Dengan adanya *internet* akan mempermudah dan mempercepat proses pengolahan data, mencari informasi dan lain-lain. Salah satu fasilitas pendukung perkembangan *internet* adalah *search engine* (mesin pencarian).

*Search engine* (mesin pencarian) yaitu kombinasi perangkat keras dan perangkat lunak komputer yang disediakan oleh perusahaan tertentu melalui *website* yang telah ditentukan. Banyak peneliti dan *survey* menunjukkan bahwa *Google* adalah *search engine* nomor satu diikuti oleh Yahoo (Abdillah, Falkner, & Hemer, 2010). *Search engine* mengumpulkan informasi dari *web* melalui program *bot* (*robot*) atau *web crawler* yang secara periodik menelusuri *web*. Proses pengambilan informasi dari *website-website* yang ada ini disebut dengan "*web scraping*".

*Web Scraping* (Turland, 2010) adalah proses pengambilan sebuah dokumen semi-terstruktur dari internet, umumnya berupa halaman-halaman *web* dalam bahasa *markup* seperti HTML atau XHTML, dan menganalisis dokumen tersebut untuk diambil data tertentu dari halaman tersebut untuk digunakan bagi kepentingan lain.

*Web scraping* sering dikenal sebagai *screen scraping*. *Web Scraping* tidak dapat dimasukkan dalam bidang *data mining* karena *data mining* menyiratkan upaya untuk memahami pola semantik atau tren dari sejumlah besar data yang telah diperoleh. Aplikasi *web scraping* (juga disebut *intelligent, automated, or autonomous agents*) hanya fokus pada cara memperoleh data melalui pengambilan dan ekstraksi data dengan ukuran data yang bervariasi.

Web scraping memiliki sejumlah langkah, sebagai berikut: 1) *Create Scraping Template*: Pembuat program mempelajari dokumen HTML dari *website* yang akan diambil informasinya untuk tag HTML yang mengapit informasi yang akan diambil, 2) *Explore Site Navigation*: Pembuat program mempelajari teknik navigasi pada *website* yang akan diambil informasinya untuk ditirukan pada aplikasi *web scraper* yang akan dibuat, 3) *Automate Navigation and Extraction*: Berdasarkan informasi yang didapat pada langkat 1 dan 2 di atas, aplikasi *web scraper* dibuat untuk mengotomatisasi pengambilan informasi dari *website* yang ditentukan, dan 4) *Extracted Data and Package History*: Informasi yang didapat dari langkah 3 disimpan dalam tabel atau tabel-tabel *database*. Cara kerjanya lihat gambar 1 (The Computer Advisor).

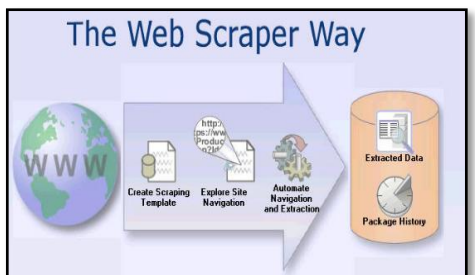

*Gambar 1. Ilustrasi Cara Kerja Web Scrapper*

Manfaat dari *web scraping* ialah agar informasi yang dikeruk lebih terfokus sehingga memudahkan dalam melakukan pencarian sesuatu, adapun cara mengembangkan teknik *web scraping* yaitu dengan cara pertama Pembuat program mempelajari dokumen HTML dari *website* yang akan diambil informasinya untuk di tag HTML tujuannya ialah untuk mengapit informasi yang diambil (*Create Scraping Template),* setelah itu pembuat program mempelajari teknik navigasi pada *website* yang akan diambil informasinya untuk ditirukan pada aplikasi *web scraping* yang akan dibuat (*Explore Site Navigation*), kemudian aplikasi *web scraping* akan mengotomatisasi informasi yang didapat dari *website* yang telah ditentukan (*Automate Navigation and Extraction*), informasi yang didapat tersebut akan disimpan ke dalam tabel basisdata (*Extracted Data and Package History*) (Juliasari & Sitompul, 2012).

Sejumlah penelitian terkait *web srcapping*, antara lain: 1) Aplikasi *Search Engine Paper* Karya Ilmiah Berbasis Web (Darmadi, Intan, & Lim, 2006), 2) Penghasil konten otomatis halaman *web* (Utomo, 2012), 3) Aplikasi *Search Engine* dengan Metode *Depth First Search* (DFS) (Juliasari & Sitompul, 2012), dan 4) *Web Scraping* pada Situs Wikipedia (Utomo, 2013).

Berdasarkan latar belakang di atas penulis tertarik untuk meneliti, merancang dan mengimplementasikan sebuah aplikasi pencarian dengan menggunakan bahasa pemrograman PHP dan *database* MySQL sebagai medium penyimpanan datanya yang akan dimanfaatkan secara spesifik untuk mengumpulkan informasi mengenai artikel pada dokumen ilmiah. Dokumen ilmiah (Abdillah, 2012) yang paling populer adalah artikel jurnal ilmiah dari suatu bidang atau topik. Apalagi saat ini, peningkatan volume literatur ilmiah yang dipublikasikan baik dalam format naskah dan juga tersedia secara elektronik (Abdillah, Falkner, & Hemer, 2011).

## 2. METODOLOGI

Waktu penelitian dimulai dari awal bulan Desember 2013 sampai dengan akhir bulan Januari 2014 (dua bulan). Penelitian penerapan teknik *web scraping* dilakukan dengan observasi pada sejumlah portal gratis, sebagai berikut: 1) Portal Garuda, 2) Portal *Indonesian Scientific Journal Database* (ISJD), dan 3) Portal Google Cendekia (*Google Scholar*).

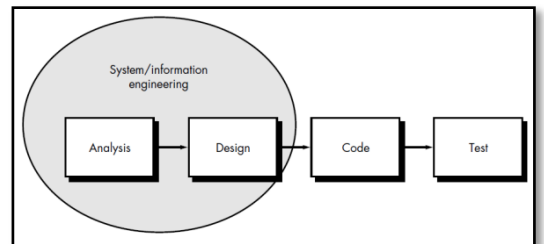

*Gambar 2. Linear Sequential Model*

Metode pengembangan sistem yang digunakan adalah *linear sequential model* (Pressman, 2001) model atau *classic life cycle* atau *waterfall model,* yang melibatkan fase-fase: 1) *Analysis*, 2) *Design*, 3) *Coding*, dan 4) *Testing*.

### 2.1 Analisis

Kegiatan analisis melibatkan 4 aktivitas, yang terdiri atas: 1) *Initiating the Process*, 2) *Facilitated Application Specification Techniques*, 3) *Quality function deployment* (QFD): a) *Normal requirements*, b) *Expected requirements*, c) *Exciting requirements*, dan 4) *Use-Cases*.

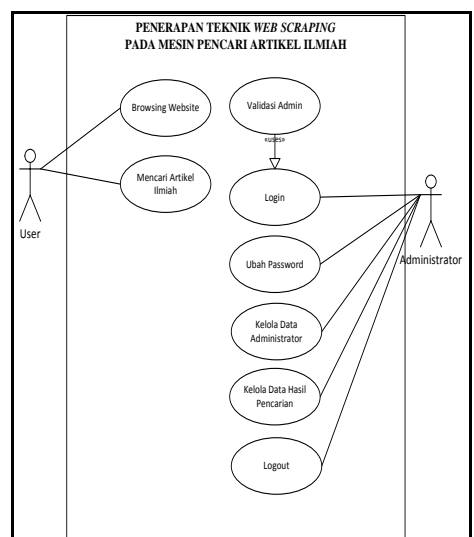

*Gambar 3. Use-Case Diagram*

Pada *use-case* (gambar 3) memperlihatkan uraian kegiatan yang melibatkan *administrator* dan *user*. *Administrator* mengelola data admin (jika diperlukan) sekaligus mengelola data hasil pencarian. Sementara *user* hanya diperbolehkan untuk melakukan pencarian (*searching*).

## 2.2 Perancangan Basisdata (*Database Design*)

Pada bagian desain, penulis fokus pada desain tabel data *scrape* (tabel 1). Desain *database* merupakan salah satu cara untuk merancang dan membangun sistem, dalam hal ini *web scraping*.

Table 1. Tabel *Data_Scrape*

| No | Nama Field | Type | Notes |
|---|---|---|---|
| 01 | id | int(4) unsigned zerofill | ID Pencarian (*primary Key*) |
| 02 | website | varchar(200) | Alamat *website* yang di-*scrape* |
| 03 | keyword | varchar(400) | *Keyword* yang dicari |
| 04 | hasil | text | Hasil Pencarian |
| 06 | file download | varchar(400) | File yang di-*download* (Jika Ada) |
| 07 | tgl jam_update | timestamp | Tanggal dan Jam *update* |

## 2.3 Pembuatan Kode (*Coding*)

Pada fase ini rancangan yang telah dibuat diterjemahkan ke dalam bentuk bahasa pemrograman dalam hal ini menggunakan bahasa pemrograman PHP. Koding berfungsi untuk menjalankan aplikasi dengan logika algoritma yang telah di terjemahkan, agar aplikasi berjalan sesuai dengan harapan maka koding harus sesuai dengan alur rancangan.

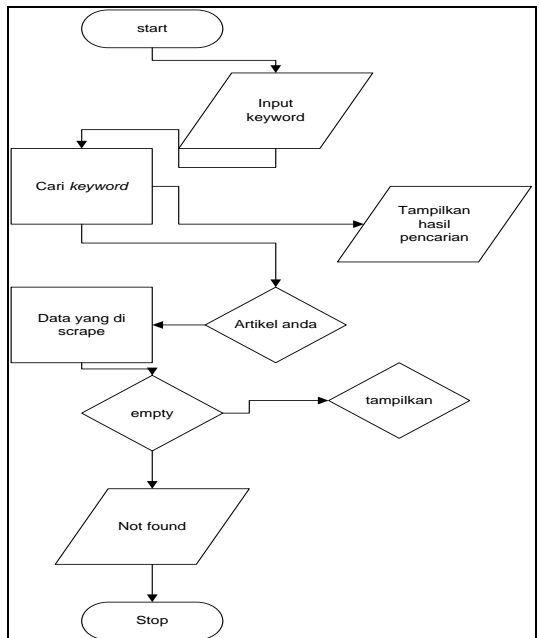

*Gambar 4. Logika Pencarian Web Scraping*

Aliran proses pencarian artikel pada *website* dengan *web scraping* diperlihatkan pada gambar 4. Variabel *keyword* dimasukkan sebagai dasar pencarian artikel, jika artikel ditemukan di dalam *database* maka artikel akan ditampilkan, namun jika artikel tidak ditemukan maka dilakukan proses *web scraping* untuk kata kunci yang dimasukan. Jika *web scraping* berhasil menemukan artikel maka akan ditampilkan ke halaman *web*, apabila tidak maka akan ditampilkan pesan *error Not Found* (Utomo, 2012).

Berikut langkah-langkah untuk mencari data *scrape* dari *website*: 1) Masukkan pencarian, 2) aplikasi akan melakukan pencarian, dan 3) Jika data di temukan > data akan di tampilkan, jika tidak, maka akan di tampilkan pesan data tidak di temukan. Cuplikasi koding pencarian dapat dilihat pada gambar 5.

```
<?php
if (isset($_POST['txtcari']) &&
$_POST['txtcari']!='' ) {
//echo trim($html);
$hasil = json_decode($html);
//print_r($hasil);
?>
```
*Gambar 5. Cuplikan Koding*

## 2.4 Pengujian (*Testing*)

Pada fase ini apilkasi *web scraping* yang telah berhasil dibuat programnya dilakukan pengujian atau *testing* untuk mencari kesalahan *coding* dan kesalahan logika, pengujian ini menggunakan *black-box testing* dan pengujian dengan pendekatan *top-down* (Sobri & Abdillah, 2013).

Table 2. Pengujian *Top Down*

| Home Page | | | |
|---|---|---|---|
| P | I | G | L |
| P1 | I1 | G1 | L1 |

Tabel 2 menampilkan pengujian aplikasi *web scraping* dengan pendekatan *top-down,* aplikasi dibagi menjadi empat modul utama (P, I, G, L). pengujian dilakukan dimulai dari sisi sebelah kiri (modul P). Apabila modul telah berjalan dengan baik maka pengujian berpindah kearah kanan, sampai dengan modul L.

Setelah semua modul melewati fase pengujian, didapatilah sebuah aplikasi yang berjalan dengan sempurna. Aplikasi ini mempunyai *output* berupa hasil pencarian, dan nantinya hasil pencarian tersebut akan disimpan pada *database*, hasil data *scrape* dapat dilihat di halaman *administrator*. Adapun pembahasan yang dimulai dari pembahasan menu *homepage*, menu pencarian artikel pada portal garuda, ISJD, *Google Scholar*, dan *login administrator*.

## 3. HASIL DAN PEMBAHASAN

Hasil dari penelitian ini adalah berupa Aplikasi *web* yang mengimplementasi teknik *web scraping* pada mesin pencari artikel ilmiah dengan menggunakan bahasa pemrograman PHP dan hasil pencariannya ditampung ke dalam tabel menggunakan *database* MySQL.

### 3.1 Home Page

Pada saat aplikasi *web scraping* ini dijalankan, aplikasi *web* akan menampilkan halaman depan (*homepage*) yang berisi tentang penjelasan mengenai *web scraping* dan juga penjelasan mengenai artikel ilmiah, penjelasan ini bertujuan untuk membuat pengguna atau *user* dapat mengetahui mengenai *web scraping*. Berikut adalah gambar dari *home page* aplikasi *web scraping*.

### 3.2 Pencarian Artikel Ilmiah pada Portal Garuda

Halaman ini menyediakan sarana atau tempat pencarian artikel ilmiah pada portal Garuda, pengguna dapat melakukan pencarian dengan mengetikan *keyword* pada kolom pencarian yang telah disediakan oleh aplikasi ini, kemudian pengguna menekan tombol cari, setelah itu aplikasi akan melakukan pencarian pada *web* yang telah di *scrape* (portal garuda), dan jika memang data yang dicari ada pada *website* maka hasilnya akan tampil (gambar 7).

### 3.3 Pencarian Artikel Ilmiah pada Potal ISJD

Halaman ini menyediakan sarana atau tempat pencarian artikel ilmiah pada ISJD *Indonesian Scientific Journal Database*, pengguna atau *user* dapat melakukan pencarian dengan cara mengetikan *keyword* pada kolom pencarian yang telah disediakan oleh aplikasi ini kemudian pengguna menekan tombol pilihan pencarian berdasarkan apa yang akan di cari di sini ada tiga kategori pilihan yaitu berdasarkan " judul, pengarang, dan *keyword* "kategori pilihan ini menirukan pencarian pada *website* ISJD, setelah itu tekan tombol cari, selanjutnya aplikasi akan melakukan pencarian pada *web* yang telah di *scrape* dalam hal ini *website Indonesian Scientific Journal Database* (ISJD), dan jika data yang dicari ada pada *website* maka akan ditampilkan namun jika tidak ada maka akan di tampilkan pesan data tidak ditemukan, untuk lebih jelasnya dapat dilihat pada gambar 8.

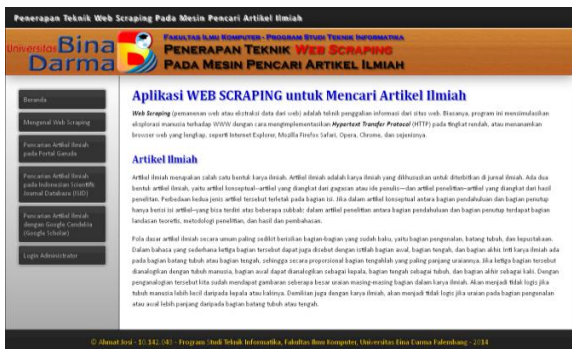
Gambar 6. Home Page

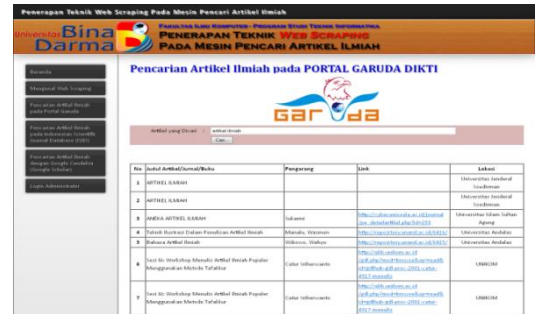
*Gambar 7. Pencarian Artikel pada Portal Garuda*

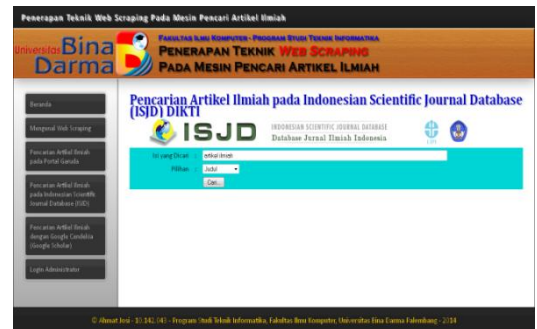
*Gambar 8. Pencarian Artikel pada Portal ISJD*

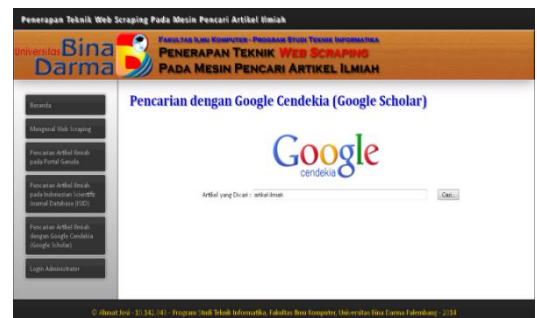
*Gambar 9. Pencarian Artikel pada Portal Google Scholar*

*Gambar 10. Hasil Scraping dari Portal Google Scholar*

### 3.4 Pencarian Artikel Ilmiah pada Portal *Google Scholar*

Halaman ini menyediakan sarana pencarian artikel ilmiah pada portal Google Cendekia (*Google Scholar*). Kelebihan dari portal *Google Scholar* adalah dibanding portal Garuda dan ISJD adalah: 1) Jumlah artikel/jurnal/buku sangat banyak dibanding portal Garuda atau ISJD, dan 2) Jika *link* yang ditemukan menyediakan fasilitas *download* artikel dalam format PDF/Word/PowerPoint/Postscript dan lain-lain, link *download* tersebut langsung terlihat. Tampilan halaman pencarian dengan Google Cendekia. Adapun data hasil dari *scraping* yang tersimpan di dalam *database* dapat dilihat pada gambar 10.

Data tersebut merupakan data hasil dari *scrape* terhadap beberapa *websites*, hasil *scrape* tersebut sudah tersimpan di dalam *database* dan dapat di akses melalui halaman admin dengan cara *login* menggunakan *password admin* dan kemudian masuk ke halaman data *scraping*.

### 3.5 Aspek Legal *Web Scrapping*

*Data scraping* biasanya mengumpulkan data dari *screen outputs* atau mengekstrak data dari kode *HyperText Markup Language* ("HTML") yang paling sering ditampilkan oleh sebagian besar *websites* (Lindenberg). Sehingga aplikasi tidak mengambil dari selain dari yang disediakan/dihasilkan oleh *interface website* yang di-*scrape*.

Algoritma akan menganalisis analisis konten halaman situs (Bakaev & Avdeenko, 2014). *Web scraping* yang diterapkan pada penelitian ini hanya berhubungan dengan informasi yang berhubungan dengan *metadata* atau informasi terkait *bibliography* dari suatu artikel ilmiah. Sehingga apabila *link* lokasi pdf file yang ter-*scrap* merupakan *open access link*, maka *user* dapat men-*download* pdf artikel tersebut. Namun, apabila *link* tersebut bersifat *close source*, tentu pihak *web host* akan mem-*block access* hanya kepada yang memiliki otoritasnya, misalnya dengan memasukkan *id* dan *pasword*. Sehingga aspek legal dari aplikasi yang dibahas pada artikel ini tidak melanggar pihak manapun. Selain itu, penerapan pada artikel ini adalah pada portal-portal yang menyediakan layanan gratis, seperti: 1) Portal Garuda, 2) *ISJD*, dan 3) *Google Scholar*.

Selain itu, *scrapers* pada artikel ini bersifat "mutual benefit" yang dapat membantu "scraped websites" (Hirschey, 2014) mendesiminasikan artikel ilmiah mereka, karena tujuan utama dari situs artikel ilmiah adalah menjadikan koleksi mereka diakses oleh sebanyak mungkin golongan yang membutuhkannya.

Selanjutnya, pihak ketiga (*scrapper*) boleh beroperasi *deep-links* dengan membuat suatu *page* dengan cara tertentu sehingga pengguna *website* ditampilkan dengan informasi yang muncul seperti yang dimiliki penaut, dimana sang *linker* membenamkan suatu *hyperlink* di kodenya untuk membawa *user* ke konten data yang asli dari pemilik situs (Jennings & Yates, 2009).

### 4. SIMPULAN dan SARAN

Berdasarkan hasil implementasi aplikasi *web scraping* pada sejumlah portal *search engines* (Garuda, ISJD, *Google Scholar*), penulis menarik kesimpulan sebagai berikut:
1) Aplikasi *search engine* yang dihasilkan dengan menerapkan teknik *web scraping* ini berhasil mengekstrak informasi mengenai artikel jurnal ilmiah dari sejumlah portal akademik baik yang berasal dari Indonesia maupun luar negeri.
2) Aplikasi ini berhasil menyimpan otomatis data hasil *scraping* pada *database*.
3) Dengan adanya aplikasi ini, pengguna dapat dengan mudah untuk mengumpulkan informasi mengenai artikel/jurnal ilmiah.
4) Teknik *web scraping* merupakan suatu teknik yang sangat bermanfaat untuk mendapatkan data artikel ilmiah secara cepat dari halaman-halaman *web*.
5) *Web scraping* adalah legal selama tidak dilakukan untuk pencurian data, manipulasi informasi, dsb. Bahkan *web scraping* dapat memberikan mutual simbiosme dengan meningkatnya trafik atas sumber asli *link* yang di-*scrape*.
6) Selanjutnya, penulis tertarik untuk: a) menerapkan teknik ini ke aplikasi *web base* lainnya yang berkaitan dengan pendidikan, politik, wisata, dll., b) melakukan percobaan *scrape* ke sejumlah *web* secara bersamaan, c) pengembangan dengan menggunakan bahasa pemrograman yanag berbeda, serta d) percobaan yang melibatkan konten yang lebih banyak.

### 5. DAFTAR RUJUKAN